\documentclass{article}
\usepackage{romp31,epsfig}

\title{Shape-Changing Collisions of Coupled Bright Solitons in 
Birefringent Optical Fibers}
\author{ M.Lakshmanan, T.Kanna and R.Radhakrishnan
\\ Centre for Nonlinear Dynamics\\
Bharathidasan University\\ Tiruchirapalli 620 024\\
India\\[2ex]
 lakshman@bdu.ernet.in}
       
\begin{document}

\maketitle
\begin{abstract}
We critically review the recent progress in understanding soliton propagation
in birefringent optical fibers.    
By constructing the most general bright two-soliton solution of the 
integrable coupled nonlinear Schr{\"o}dinger equation (Manakov model) we
point out that solitons in birefringent fibers can in general change their shape
after interaction due to a change in the intensity distribution among the modes 
eventhough the total energy is conserved. However the standard shape-preserving
collision (elastic collision) property of the $(1+1)$ dimensional solitons is 
recovered when restrictions are imposed on some of the soliton parameters. As a
consequence the following further properties can be deduced using this
shape-changing collision.

(i) The exciting possibility of switching of solitons between orthogonally polarized
modes of the birefringent fiber exists.

(ii) When additional effects due to periodic rotation of birefringence axes are
considered, the shape changing collision can be used as a switch to suppress or
to enhance the periodic intensity exchange between the orthogonally polarized
modes.

(iii) For ultra short optical soliton pulse propagation in
non-Kerr media, from the governing equation an integrable system of
coupled nonlinear Schr{\"o}dinger equation with cubic-quintic
terms is identified. It admits a nonlocal Poisson bracket structure.

(iv) If we take the higher-order terms in the coupled nonlinear Schr{\"o}dinger
equation into account then their effect on the shape changing collision of solitons, during
optical pulse propagation, can be studied by using a direct perturbational
approach.

\end{abstract}

\section{Introduction}

The study of optical wave propagation in a nonlinear dispersive (dielectric)
fiber has been receiving considerable attention in recent times as the fiber
can support under suitable circumstances a stable pulse called optical soliton
[1-3]. It arises essentially due to a compensation of the effect of dispersion of 
the pulse by the nonlinear response of the medium. 

The analysis of such pulse
propagation naturally starts from the Maxwell's equations for electromagnetic
wave propagation in a dielectric medium,
\begin{equation}\label{1a}
\nabla^2\overrightarrow{E}-\frac{1}{c^2}
\frac{\partial^2 \overrightarrow{E}}
{\partial t^2}  =  -\mu_0 \frac{\partial^2 \overrightarrow{P}}
{\partial t^2},
\end{equation}
where the induced polarization  for silica fibers is
\begin{equation}
\overrightarrow{P}(\overrightarrow{r},t)  =  \epsilon_0
\left[\chi^{(1)}.\overrightarrow{E} \right.\nonumber \\
 \left. +\chi^{(3)} \vdots \overrightarrow{E}\overrightarrow{E}
\overrightarrow{E}+...\right].
\end{equation}
In equation(1), $\overrightarrow {E}$ represents the electric field, c is
the velocity of light, $\mu_0$ and $\epsilon_0$ are the permeability and
permitivity of free space respectively, $\chi^{(m)}$ is the mth order 
susceptibility tensor.

In order to analyse equation(1) it is necessary to make several simplifying
assumptions:
(i)   The nonlinear part of the induced polarization
 is treated as a small
perturbation to the linear part.
(ii)  The optical field is assumed to maintain its polarization along the fiber.
(iii) Fiber loss is assumed to be very small.
(iv)  The nonlinear response of the fiber is assumed to be instantaneous.
(v)   In a slowly varying envelope approximation for pulse propagation along the 
fiber, the electric field can be written as [4]
\begin{equation}\label{2}
\overrightarrow{E}(\overrightarrow{r},t)  =\frac{1}{2}
\overrightarrow{e}\Big[F(x,y)
E(z,t) \nonumber \\
 e^{i(K_0z-\omega_0t)}+c.c\Big ],
\end{equation}
where c.c stands for complex conjugate, $\overrightarrow e$ is the unit 
polarization vector of the light assumed to be linearly polarized, 
E is the slowly varying electric field, $F(x,y)$ is the mode distribution function in the $(x,y)$ plane, 
while $K_0$ and $\omega_0$ denote the propagation constant and central frequency
of the optical pulse respectively.

Under the above assumptions, rewriting equation(1) by using the method of separation 
of variables and introducing the coordinate system,
 $T$$=$$t$$-$$\frac{z}{v_g}$,
moving with the pulse at the group-velocity $v_g$ = 
$\frac{\partial K}{\partial \omega}$, one can obtain a wave
equation for the evolution of E
as
\begin{equation}
i\frac{\partial E}{\partial z}-\frac{K''}{2}\frac{\partial^2 E}{\partial T^2}
+\gamma_0|E|^2E = 0,
\end{equation}
where
$\gamma_0$$=$$\frac{n_2\omega_0}{cA_{eff}}$.
Here $A_{eff}$ denotes the effective core area of the single-mode fiber,
$n_2$ represents the nonlinear index coefficient and the parameter
$K''=\left(\frac{\partial^2 K}{\partial \omega^2}\right)_{\omega=\omega_0}
=\frac{1}{v_g^2}
$ accounts for the group velocity dispersion (GVD). After normalizing
equation(4) and using the transformation, $q = \left[\frac{\gamma_0T_0^2}
{|K''|}\right]^{\frac{1}{2}}E,\;\;\xi= \frac{z|K''|}{T_0^2},\;\;
\tau = \frac{T}{T_0}$ and then redefining $\xi$ as z and $\tau$ as t
 we get the ubiquitous nonlinear Schr{\"o}dinger (NLS) equation,
\begin{equation}
iq_{z} -(\mbox{sgn}\,K'')q_{tt} + 2|q|^2q = 0, 
\end{equation}
in which $T_0$ represents the width of the incident pulse, z and t are
the normalized distance and time along the direction of propagation and
q is the normalized envelope. The NLS equation admits the familiar
bright envelope (one) soliton
\begin{equation}
q(z,t) = \frac{k_{1R}exp\left[i\left(\left(k_{1R}^2-k_{1I}^2 \right)z+k_{1I}t
+\eta_{1I}^{(0)}\right)
 \right]}
{\mbox{cosh}\left[k_{1R}\left(t-k_{1I}z+\eta_{1R}^{(0)}\right)\right ]},
\end{equation}
in the anamolous dispersion region $(K''<0)$, where $k_{1R}$ 
and $k_{1I}$ give the amplitude and velocity of the soliton respectively 
and $\eta_1^{(0)}$ is an arbitrary complex parameter.
 In the normal dispersion
regime $(K''>0)$, equation(5) possesses dark soliton solutions [5].
Equation(5) is 
valid in the picosecond regime since in obtaining the above equation
the optical field is assumed to be  quasi-monochromatic. For ultra short 
pulses (width $\leq 100fs$) one has to include additional effects
such as third order dispersion, nonlinear dispersion, self induced Raman
effect, etc. Then equation(5) is for example replaced by the higher-order NLS 
(HNLS) equation,
\begin{equation}
iq_z - (\mbox{sgn}\,K'')q_{tt} + 2|q|^2q -\epsilon q_{ttt} 
-\lambda_1(|q|^2q)_t -\lambda_2|q|^2_tq =  0.
\end{equation}
Considerable attention has been paid to this HNLS equation which 
results from the delayed response of the fiber nonlinearity [6].
\section{Electromagnetic Wave Propagation in  
Birefringent Fibers and the  Coupled NLS Equation }

In general a single mode fiber can support two distinct modes of
polarization which are orthogonal to each other. This phenomenon
is known as birefringence. Among these two modes one corresponds to
the ordinary ray (O-ray) in which the refractive index of the 
medium is constant along every direction of the incident ray.
The other is the extraordinary ray (E-ray) whose refractive index
for the medium varies with the direction of the incident ray. In an
ideal fiber these two modes are degenerate, while in a real fiber 
due to the fiber nonlinearity this degeneracy is broken and the
phenomenon is known as ``modal birefringence" [1].

Thus due to the effect of birefringence and nonlinear response of
the medium there is a possibility of interaction of two
copropagating modes. As a result of this the phase of one mode not
only depends on its own intensity (Self Phase Modulation (SPM)) 
but also on the intensity of the copropagating mode (Cross Phase 
Modulation (XPM)) [1,2].

The propagation equation for such modes can be obtained again from 
the Maxwell's equations by considering the electric field $\overrightarrow E$
in the slowly varying envelope approximation as
\begin{equation}
\overrightarrow{E}(\overrightarrow{r},t)  =  \frac{1}{2}
\left [(\overrightarrow{e_1}E_1(z,t)e^{iK_{01}z}
+\overrightarrow{e_2}E_2(z,t)\right.
 \left. e^{iK_{02}z})F(x,y)e^{-i\omega_0t}+c.c\right ],
\end{equation}
where $E_1$, $E_2$ are the 
amplitudes of two polarization components and  $\overrightarrow e_1$,
$\overrightarrow e_2$ are the unit orthonormal polarization vectors and
$F(x,y)$ is the fiber mode distribution.

Proceeding in the same way as in the case of the single mode fiber
one can
obtain the following coupled system of equations for the envelopes of the
two copropagating waves,
\begin{eqnarray}
iE_{1z}+\frac{i}{v_{g1}}E_{1t}-\frac{K''}{2}E_{1tt}
+\gamma_0(|E_1|^2 
+B|E_2|^2)E_1  =  0,\nonumber\\
iE_{2z}+\frac{i}{v_{g2}}E_{2t}-\frac{K''}{2}E_{2tt}
+\gamma_0(|E_2|^2 
+B|E_1|^2)E_2  =  0,
\end{eqnarray}
where z and t represent the normalized distance and time 
along the direction of propagation, 
 $v_{g1}$ and $v_{g2}$ represent the group velocity of the two
copropagating waves $E_1$ and $E_2$ respectively and  
$B = \frac{2+2\sin^2\vartheta}{2+\cos^2\vartheta}$ is the XPM coupling
parameter ($\vartheta$ - birefringence ellipticity angle which varies
between 0 and $\frac{\pi}{2}$ ). After suitable transformations as
in the case of NLS equation we can rewrite the equation(9) 
in its normalized form as 
\begin{eqnarray}
iq_{1z}+q_{1tt}+2\mu(|q_1|^2+B|q_2|^2)q_1  =  0,\nonumber\\
iq_{2z}+q_{2tt}+2\mu(|q_2|^2+B|q_1|^2)q_2  =  0.
\end{eqnarray}
The above equation is the coupled nonlinear Schr{\"o}dinger (CNLS) equation 
which is in general nonintegrable. However for B$=$1 this reduces 
to the celebrated Manakov equations [7],
\begin{eqnarray}
iq_{1z}+q_{1tt}+2\mu(|q_1|^2+|q_2|^2)q_1  =  0,\nonumber\\
iq_{2z}+q_{2tt}+2\mu(|q_1|^2+|q_2|^2)q_2  =  0,
\end{eqnarray}
which is a completely integrable soliton system.  The Lax pair for the
Manakov system (11) can be identified as
\begin{eqnarray}
L=
\left(
\begin{array}{ccc}
-i\lambda & q_1 & q_2 \\
-q_1^* & i\lambda & 0 \\
-q_2^* & 0 & i\lambda
\end{array}
\right),
 \quad
M=
\left(
\begin{array}{ccc}
\{-2i\lambda^2 +& \{2\lambda q_1  & \{2\lambda q_2  \\
i(|q_1|^2+|q_2|^2)\} & +iq_{1t}\} &+iq_{2t}\} \\
\\
-2\lambda q_1^*+iq_{1t}^*& \{2i\lambda^2   & -iq_{1}^*q_2  \\
 &-i|q_1|^2\} &
\\
-2\lambda q_2^*+iq_{2t}^*& -iq_{1}q_2^*  &\{2i\lambda^2    \\
 & & -i|q_2|^2\}\\
\end{array}
\right),
\end{eqnarray}
such that
\begin{equation}
L_z - M_t + [L,M] = 0,
\end{equation}
which is equivalent to the Manakov equations (11). The existence of 
infinite number of involutive integrals of motion confirms the
integrability of the Manakov system (11).
\section{Bilinearization and Bright Two-Soliton Solution}
In recent years to study the solution properties of the
integrable systems several effective 
tools have been developed which include inverse scattering transform method,
Hirota's bilinearization method, B{\"a}cklund transformation method, geometrical
methods and so on. In this section by applying Hirota's technique we point
out that the most general bright one-soliton and two-soliton solutions for the
Manakov system (11) can be obtained [8].

Considering equation (11) and by making the following bilinear transformation
$q_1 = \frac{g}{f}$, $q_2 = \frac{h}{f}$, where g(z,t), h(z,t) are complex 
functions while f(z,t) is a real function, the following bilinear equations 
can be obtained,
\begin{equation}
(iD_z+D_t^2)g.f= 0,\;\;
(iD_z+D_t^2)h.f = 0,\;\;
D_t^2f.f= 2\mu(gg^*+hh^*),
\end{equation}
where $D_z$ and $D_t$ are Hirota's bilinear operators.
The above set of equations can be solved by making the following power
series exapansion to g, h and f: 
\begin{equation}
g  =  \lambda g_1 + \lambda^3 g_3 +...,\;\; h = \lambda h_1 + \lambda^3 h_3 +...,\;\;
f  = 1 +\lambda^2 f_2 +\lambda^4 f_4+...,
\end{equation}
where $\lambda$ is the formal expansion parameter.
The resulting equations, after collecting the terms with the same power in $\lambda$, 
can be solved to obtain the forms of g, h and f. In order to get the
one-soliton solution the power series expansions for g, h and f are terminated
as follows, $g$$=$$\lambda g_1$, $h$$=$$\lambda h_1$ and  $f$ $=$ $1$ $+
$$\lambda^2 f_2$.
After following the procedure as mentioned before, the bright one-soliton
solution is obtained as
\begin{eqnarray}
\left(
\begin{array}{c}
q_1\\
q_2 
\end{array}
\right)
 = 
\left(
\begin{array}{c}
\alpha\\
\beta
\end{array}
\right)\frac{e^{\eta_1}}{1+e^{\eta_1+\eta_1^*+R}}\;\;
 = 
\left(
\begin{array}{c}
A_1 \\
A_2
\end{array}
\right)
\frac{k_{1R}e^{i\eta_{1I}}}{\mbox{cosh}\,(\eta_{1R}+\frac{R}{2})}, 
\end{eqnarray}
where $\;$
$
\eta_j =k_j(t+ik_jx)+\eta_j^{(0)}
 ,\;\; j=1,\;\; 
A_1 = \frac{\alpha}{\Delta},\;\;A_2 = \frac{\beta}{\Delta},\;\;
\Delta = \sqrt{\mu(|\alpha|^2+|\beta|^2)},
$$\;\;$
$e^R=\frac{\mu(|\alpha|^2+|\beta|^2)}{(k_1+k_1^*)^2}$,$\;\;$
$\alpha$,$\;\;$$ \beta$, $k_j$ and $\eta_j^{(0)}$ are complex parameters.                                                                                                      	
Here $\sqrt{\mu} (A_1,A_2)$ represents the unit polarization vector,
$k_{1R}$ and $k_{1I}$ give the amplitude and velocity of the Manakov one
soliton respectively. 

For obtaining the bright two-soliton solution, the series is terminated
as $g =  \lambda g_1 +\lambda ^3g_3 ,\;\; h =
 \lambda h_1+\lambda^3 h_3 $ and
$f  = 1 +\lambda^2 f_2 +\lambda^4 f_4.$
After proceeding in a similar fashion as in the case of the 
one-soliton solution, the following
bright two-soliton solution with six  arbitrary complex 
parameters $k_1$, $k_2$, $\alpha_1$, $\beta_1$,                                                                                                                                                                                                                                                                                                                                                                                                                                                                                               
$\alpha_2$ and $\beta_2$ can be obtained,
\begin{eqnarray}
q_1&=&\frac{\alpha_1e^{\eta_1}+\alpha_2e^{\eta_2}
+e^{\eta_1+\eta_1^*+\eta_2+\delta_1}+e^{\eta_1+\eta_2+\eta_2^*
+\delta_2}}
{D},\nonumber\\
q_2 &= & \frac{\beta_1e^{\eta_1}+\beta_2e^{\eta_2}
+e^{\eta_1+\eta_1^*+\eta_2+\delta_1'}+e^{\eta_1+\eta_2+\eta_2^*
+\delta_2'}}
{D},
\end{eqnarray}
where
\begin{eqnarray}
D &=& \displaystyle 1+e^{\eta_1+\eta_1^*+R_1}+e^{\eta_1+\eta_2^*+\delta_0}
 +e^{\eta_1^*+\eta_2+\delta_0^*}+e^{\eta_2+\eta_2^*+R_2}
+e^{\eta_1+\eta_1^*+\eta_2+\eta_2^*+R_3},\nonumber\\
e^{\delta_0} &=& \frac{\kappa_{12}}{k_1+k_2^*},\;\;
e^{R_1} = \frac{\kappa_{11}}{k_1+k_1^*},\;\;\;\;
e^{R_2}=  \frac{\kappa_{22}}{k_2+k_2^*},\nonumber\\
e^{\delta_1}&=&\frac{k_1-k_2}{(k_1+k_1^*)(k_1^*+k_2)}
(\alpha_1\kappa_{21}-\alpha_2\kappa_{11}),\;\;
e^{\delta_2}=\frac{k_2-k_1}{(k_2+k_2^*)(k_1+k_2^*)}
(\alpha_2\kappa_{12}-\alpha_1\kappa_{22}),\nonumber\\
e^{\delta_1^{'}}&=& \frac{k_1-k_2}{(k_1+k_1^*)(k_1^*+k_2)}
(\beta_1\kappa_{21}-\beta_2\kappa_{11}),\;\;
e^{\delta_2^{'}}= \frac{k_2-k_1}{(k_2+k_2^*)(k_1+k_2^*)}
(\beta_2\kappa_{12}-\beta_1\kappa_{22}),\nonumber\\
e^{R_3}&=&  \frac{|k_1-k_2|^2}{(k_1+k_1^*)(k_2+k_2^*)|k_1+k_2^*|^2}
 (\kappa_{11}\kappa_{22}-\kappa_{12}\kappa_{21})
\;\;\mbox{and}\;\;
\kappa_{ij} = \frac{\mu(\alpha_i\alpha_j^*+\beta_i\beta_j^*)}{k_i+k_j^*}.\nonumber
\end{eqnarray}
The above most general bright two-soliton solution corresponds to a 
shape changing (inelastic)
collision of two bright solitons which will be explained in the following
sections.
\section{Inelastic Collision and Switching of Bright Solitons in the
Manakov Model}
The collision property of bright solitons can be revealed
by analysing the asymptotic form of the two-soliton solution [8].
Without loss of generality we assume that 
$
 k_{iR} > 0 \;\;\mbox{and}\;\;
k_{1I} > k_{2I},
$ where i=1,2, which corresponds to a head-on 
collision. One can easily check that 
asymptotically the two-soliton solution becomes two well separated 
solitons $S_1$ and $S_2$. Thus for the above condition, asymptotically 
the $\eta_{iR}$'s for the two-solitons become as
(i)$\eta_{1R}\sim 0$, $\eta_{2R} \rightarrow \pm\infty$ as 
$z \rightarrow \pm \infty$$\;$and$\;$
(ii)$\eta_{2R}\sim 0$, $\eta_{1R} \rightarrow\mp\infty$ as $z \rightarrow \pm
\infty.$$\;\;$
This leads to the following asymptotic forms for the two-soliton solution.\\
(i)\underline{Limit z $\rightarrow -\infty$:}
(a) $S_1$ ($\eta_{1R}\sim 0,\;\; \eta_{2R}\rightarrow -\infty):$ 
\begin{eqnarray}
\left(
\begin{array}{c}
q_1\\
q_2 
\end{array}
\right)
\rightarrow
\left(
\begin{array}{c}
\alpha_1\\
\beta_1
\end{array}
\right)\frac{e^{\eta_1}}{1+e^{\eta_1+\eta_1^*+R_1}} 
= 
\left(
\begin{array}{c}
A_1^{1-} \\
A_2^{1-}
\end{array}
\right)
k_{1R}e^{i\eta_{1I}}{\mbox{sech}\,\left(\eta_{1R}+\frac{R_1}{2}\right)}, 
\end{eqnarray}
where $(A_1^{1-},A_2^{1-}) = [\mu(\alpha_1\alpha_1^*+
\beta_1\beta_1^*)]^{-\frac{1}{2}}(\alpha_1,\beta_1)$. Here superscript
1- denotes $S_1$ at the limit $z\rightarrow -\infty$ and subscripts 1
and 2 refer to the modes $q_1$ and $q_2$. 

(b) $S_2$ ($\eta_{2R}\sim 0,\;\; \eta_{1R}\rightarrow\infty)$: 
\begin{equation}
\left(
\begin{array}{c}
q_1\\
q_2 
\end{array}
\right)
 \rightarrow 
 \left(
\begin{array}{c}
 e^{\delta_1 - R_1}\\
e^{\delta_1' - R_1}
\end{array}
\right)
\frac{e^{\eta_2}}{1+e^{\eta_2+\eta_2^*+R_3-R_1}} 
 = 
\left(
\begin{array}{c}
A_1^{2-} \\
A_2^{2-}
\end{array}
\right)
k_{2R}e^{i\eta_{2I}}{\mbox{sech}\,\left (\eta_{2R}+\frac{(R_3 - R_1)}{2}\right)},
\end{equation} 
where 
$
(A_1^{2-},A_2^{2-})=(\frac{a_1}{a_1^*})\; c \;[\mu(\alpha_2
\alpha_2^*+\beta_2\beta_2^*)]^{-\frac{1}{2}}\left[
(\alpha_1,\beta_1)\kappa_{11}^{-1}-(\alpha_2,\beta_2)
\kappa_{21}^{-1}\right]
$ in which $a_1 =(k_1+k_2^*)\left[(k_1-k_2)(\alpha_1^*\alpha_2
+\beta_1^*\beta_2)\right]^{\frac{1}{2}}$$\;\;$and$\;\;$ $c=\left[
\frac{1}{|\kappa_{12}|^2}-\frac{1}{\kappa_{11}\kappa_{22}}\right]^{-\frac{1}{2}}.$

(ii)\underline{Limit $z\rightarrow \infty$:}
(a) $S_1$ ($\eta_{1R}$ $\sim$0,$\;\;$ $\eta_{2R}$ $\rightarrow$ $\infty$):
\begin{equation}
\left(
\begin{array}{c}
q_1\\
q_2 
\end{array}
\right)
 \rightarrow
\left(
\begin{array}{c}
 e^{\delta_2 - R_2}\\
e^{\delta_2' - R_2}
\end{array}
\right) 
\frac{e^{\eta_1}}{1+e^{\eta_1+\eta_1^*+R_3-R_2}} 
 = 
\left(
\begin{array}{c}
A_1^{1+} \\
A_2^{1+}
\end{array}
\right)
k_{1R}e^{i\eta_{1I}} {\mbox{sech}\,\left(\eta_{1R}+\frac{(R_3 - R_2)}{2}\right)}, 
\end{equation}
where 
$
(A_1^{1+},A_2^{1+})=(\frac{a_2}{a_2^*})\; c \;[\mu(\alpha_1
\alpha_1^*+\beta_1\beta_1^*)]^{-\frac{1}{2}}\left[
(\alpha_1,\beta_1)\kappa_{12}^{-1}-(\alpha_2,\beta_2)
\kappa_{22}^{-1}\right]
$ in which $a_2 =(k_2+k_1^*)\left[(k_1-k_2)(\alpha_1\alpha_2^*
+\beta_1\beta_2^*)\right]^{\frac{1}{2}}.$

(b) $S_2$ ($\eta_{2R}\sim 0,\;\; \eta_{1R}\rightarrow -\infty)$: 
\begin{eqnarray}
\left(
\begin{array}{c}
q_1\\
q_2 
\end{array}
\right)
\rightarrow
\left(
\begin{array}{c}
\alpha_2\\
\beta_2
\end{array}
\right)\frac{e^{\eta_2}}{1+e^{\eta_2+\eta_2^*+R_2}} 
= 
\left(
\begin{array}{c}
A_1^{2+} \\
A_2^{2+}
\end{array}
\right)
k_{2R}e^{i\eta_{2I}}{\mbox{sech}\,\left(\eta_{2R}+\frac{R_2}{2}\right)}, 
\end{eqnarray}
where $(A_1^{2+},A_2^{2+}) = [\mu(\alpha_2\alpha_2^*+
\beta_2\beta_2^*)]^{-\frac{1}{2}}(\alpha_2,\beta_2)$.

In the above set of equations, the solitons before interaction are
given by equations (18) and (19) and the solitons after interaction 
are given by equations (20) and (21). We observe from the above
mentioned equations (18-21) that due to the interaction between
two copropagating solitons $S_1$ and $S_2$, their amplitudes
change from $(A_1^{1-}k_{1R},A_2^{1-}k_{1R})$ and
 $(A_1^{2-}k_{2R},A_2^{2-}k_{2R})$ to  $(A_1^{1+}k_{1R},
A_2^{1+}k_{1R})$ and  $(A_1^{2+}k_{2R},A_2^{2+}k_{2R})$ respectively,
in addition to a change in their phases by an amount $\frac{
R_3-R_1-R_2}{2}$ and  $\frac{-R_3+R_1+R_2}{2}$ respectively. The 
interesting behaviour which should be noted in this collision is that
though the amplitude and phase of each soliton change during 
collision, the total intensity of each soliton is conserved, ie.,
$|A_1^{n\pm}|^2+|A_2^{n\pm}|^2 = \frac{1}{\mu}$,
where n=1,2 represent the solitons $S_1$ and $S_2$ respectively. Thus in such a
collision there is a change in the distribution of 
intensity among the two component fields keeping the total intensity
conserved. This is shown in Fig(1), where a head-on collision of two
solitons is pictured for the parameter values, $k_1=1+i$, $k_2=2-i$,
$\beta_i=1$, $\alpha_1=1$ and $\alpha_2=\frac{(39+i80)}{89}$. Here 
initially the time profiles of the two-solitons are evenly split between the two
components $q_1$ and $q_2$. At the large positive z end the profile 
of the $S_1$ soliton is completely suppressed in the $q_1$ component while it is 
enhanced in the $q_2$ component. Noticeable changes in $S_2$ soliton also take 
place. 
\begin{figure}[h]
\centerline{\epsfig{figure=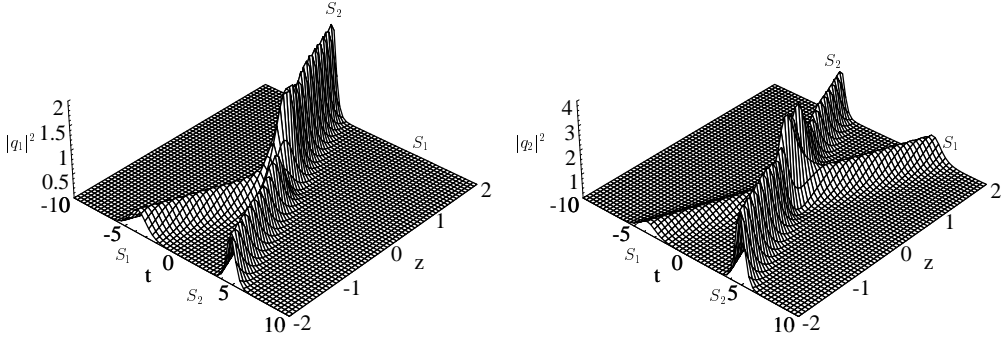, width=\linewidth}}
\caption{Intensity profiles $ |q_{1}|^2 $ and $ |q_{2}|^2 $ of the
head-on collision of the two-soliton solution of the Manakov model.
}
\label{fig1}
\end{figure}

However we can regain the elastic collision for $\alpha_2=1$ [8].
Thus for the special case $\alpha_1:\alpha_2 = \beta_1:\beta_2$
the  standard elastic collision nature of the soliton can be obtained. The
above analysis shows that the (1+1) dimensional soliton system given
by equation(11) exhibits a novel type of shape changing collision
not seen in any other (1+1) dimensional evolution equation which
led us to identify the exciting possibility of switching of solitons
between modes by changing the phase [8]. This novel type of
interaction led Jakubowski et al. to suggest a method for implementing 
computation in a bulk nonlinear medium without interconnecting 
discrete components [9].  
\section{Periodically Twisted Birefringent Fibers and Soliton
Interaction}
Soliton propagation in a periodically twisted birefringent fiber
has gained considerable attention recently [1,10]. The effect of periodic
twist on the soliton propagation can be described by the coupled
wave equations of the form
\begin{eqnarray}
iq_{1z}+q_{1tt}+ \varrho q_1 + \sigma q_2 +  
2\mu(|q_1|^2+|q_2|^2)q_1 & = & 0, \nonumber\\
iq_{2z}+q_{2tt}-\varrho q_2 + \sigma  q_1
+2\mu(|q_1|^2+|q_2|^2)q_2 & = & 0, 
\end{eqnarray}
where $ \sigma  $ and $\varrho$ are the 
normalized linear coupling
constants caused by the periodic twist of the
birefringence axes and the phase-velocity mismatch
from resonance respectively. It can be easily verified that the
transformation,
$
q_1 = \cos\theta e^{i\Gamma z}q_{1M} -\sin
\theta e^{-i\Gamma z}q_{2M}, \;\;
q_2 = \sin\theta e^{i\Gamma z}q_{1M} +\cos
\theta e^{-i\Gamma z}q_{2M},
$
where the subscript M refers to the Manakov model and 
$\Gamma = \sqrt{\varrho^2 + \sigma ^2} $ and 
$\theta = \frac{1}{2}tan^{-1}(\frac{\sigma }{\varrho})$, reduces equation(22) to the
coupled system given by equation(11). Hence by making use 
of the solutions (16) and (17) we obtain the one-soliton and two-soliton
solutions of equation(22) respectively as 
\begin{eqnarray}
q_1&=& \left \{\cos\theta
e^{i\Gamma z}A_{1M} -\sin
\theta e^{-i\Gamma z}A_{2M}\right \}q_c,\nonumber\\
q_2& =& \left\{\sin\theta
e^{i\Gamma z}A_{1M} +\cos
\theta e^{-i\Gamma z}A_{2M}\right\}q_c, 
\end{eqnarray}
where $ q_c = k_{1R}\exp(i\eta_{1I})\mbox{sech}\,(\eta_{1R}+\frac{R}{2}) $
and
\begin{eqnarray}
q_1&=&\frac{\left [\cos\theta  e^{i\Gamma z} \alpha_1
	-\sin\theta  e^{-i\Gamma z} \beta_1
	\right]e^{\eta_1} 
 +\left [\cos\theta e^{i\Gamma z} \alpha_2 
-\sin\theta e^{-i\Gamma z} \beta_2\right ]
e^{\eta_2}}{D}\nonumber\\
& & +\frac{\left [\cos\theta e^{i\Gamma z + \delta_1}
-\sin\theta e^{-i\Gamma z +\delta_1'}\right ]
e^{\eta_1+\eta_1^*+\eta_2}}{D}\nonumber\\
& &+ \frac{\left [\cos\theta e^{i\Gamma z + \delta_2}
-\sin\theta e^{-i\Gamma z +\delta_2'}\right ]
e^{\eta_1+\eta_2+\eta_2^*}}{D},
\end{eqnarray}
where all the parameters in (23) and (24) as well as the quantity D
are defined in the previous section
and $q_2$ can be obtained by just replacing
$\cos\theta$ 
by $\sin\theta$ and $\sin\theta$ 
by $-\cos\theta$  
in the above equation.
The soliton interaction for this system can be studied by carrying out
the asymptotic analysis as before.

Then the form of $S_1$ and $S_2$ before interaction 
($z\rightarrow -\infty$) is given by
\[
(q_1,q_2)=\left([\cos\theta e^{i\Gamma z} A_{1}^{1-}-
\sin\theta e^{-i\Gamma z}
A_{2}^{1-}],
[\sin\theta e^{i\Gamma z} A_{1}^{1-}
+\cos\theta e^{-i\Gamma z} A_{2}^{1-}]\right)q^{1-}
\]
and 
\[
(q_1,q_2)=\left([\cos\theta e^{i\Gamma z} A_{1}^{2-}-
\sin\theta e^{-i\Gamma z}
A_{2}^{2-}],
[\sin\theta e^{i\Gamma z} A_{1}^{2-}
+\cos\theta e^{-i\Gamma z} A_{2}^{2-}]\right)q^{2-}
\]
respectively, where $q^{1-} = k_{1R}exp(i\eta_{1I})\mbox{sech}(\eta_{1R}+\frac{R_1}{2})$,
$q^{2-}=k_{2R}exp(i\eta_{2I})\mbox{sech}(\eta_{2R}+\frac{(R_3-R_1)}{2})
$ and the polarization vectors $A_{i}^{j-},$ j=1,2 are the same as in 
equations(18,19). 
\begin{figure}[h]
\centerline{\epsfig{figure=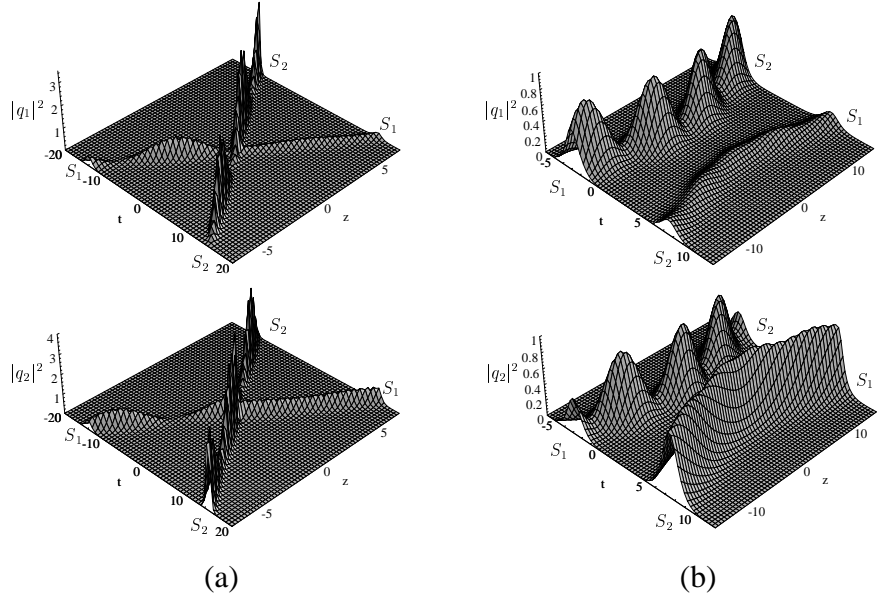, width=0.9\linewidth}}
\caption{ (a) Evolution of the intensity profiles $|q_1|^2$ and
$|q_2|^2$ showing the suppression in switching between the two modes of
$S_1$ soliton in a twisted birefringent fiber. (b) Evolution of the intensity
profiles $|q_1|^2$ and $|q_2|^2$ showing the suppression in switching of $S_1$ 
soliton and enhancement in $S_2$ soliton in a twisted birefringent fiber.}
\label{fig2}
\end{figure}

Proceeding in a similar fashion the form of $S_1$ and $ S_2$ after interaction
(limit z$\rightarrow \infty$) can be obtained as
 \[
(q_1,q_2)=\left([\cos\theta e^{i\Gamma z} A_{1}^{1+}-
\sin\theta e^{-i\Gamma z}
A_{2}^{1+}],
[\sin\theta e^{i\Gamma z} A_{1}^{1+}
+\cos\theta e^{-i\Gamma z} A_{2}^{1+}]\right)q^{1+}
\]
and 

\[
(q_1,q_2)=\left([\cos\theta e^{i\Gamma z} A_{1}^{2+}-
\sin\theta e^{-i\Gamma z} A_{2}^{2+}],  
 [\sin\theta e^{i\Gamma z} A_{1}^{2+}
+\cos\theta e^{-i\Gamma z} A_{2}^{2+}]\right)q^{2+}
\] respectively, where $q^{1+} = k_{1R}exp(i\eta_{1I})\mbox{sech}\left(\eta_{1R}+
\frac{(R_3-R_1)}{2}\right)$,
$q^{2+}=k_{2R}exp(i\eta_{2I})\mbox{sech}(\eta_{2R}+\frac{R_2}{2})
$ and the polarization vectors $A_i^{j+}$,i,j=1,2 are defined in section-4.  

In order to facilitate the understanding of the above behaviour with 
reference to the optical soliton switching between the orthogonally polarized
modes, it is convinient to obtain the oscillating part of the intensities 
associated with the above asymptotic forms as
$\left | \frac{q_l(z,t)}{q^{n\mp}(z,t)} \right |^2 = |A_{l}^{n\mp}|^2 \cos^2
\theta
+|A_{j}^{n\mp}|^2 \sin^2 \theta
 + (-1)^l|A_{l}^{n\mp}||A_{j}^{n\mp}|
\sin2\theta\cos(2\Gamma z
+\phi^{n\mp}),\;
l,j=1,2,$ where $(l \neq j),$
$ \phi^{n\mp} = \displaystyle\tan^{-1}\left(\frac{A_{1I}^{n\mp}}{A_{1R}^{n\mp}}\right)-
\tan^{-1}\left(\frac{A_{2I}^{n\mp}}{A_{2R}^{n\mp}}\right).
$ Now let us analyse how the presence of the oscillatory term in the 
above expression and the changes in $A_{j}^{n\pm}$ affect the switching
dynamics of the solitons $S_1$ and $S_2$.

\underline{case(i)}: For all the $A_{j}^{n\pm}$'s are non zero, there is a periodic
intensity switching which is always present in  both the solitons and in both
the components before as well as after the interaction.

\underline{case(ii)}: If any one of the $A_{j}^{n\pm}$'s is zero, and others are nonzero
then soliton interaction suppresses or enhances switching dynamics. This is 
illustrated in Fig(2a), for the parameter values $k_1=1+i$, $k_2=2-i$,
$\alpha_1=\beta_1=\beta_2=1$, $\alpha_2=\frac{(39+i80)}{89}$, $\rho=0.25$
and $\sigma =0.5$. Here we observe that for $|A_1^{1+}|\sim 0$ the switching
in the intensity of soliton $S_1$ is fully suppressed in both the modes $q_1$ and 
$q_2$.

\underline{case(iii)}: If any two of the $A_{j}^{n\pm}$'s are zero then the soliton 
interaction suppresses and enhances switching dynamics. This can be verified
from the Fig(2b), for the chosen parameters, $k_1=1+i(0.1)$, $k_2=1-i(0.1)$,
$\alpha_1=0.86+i(0.5)$, $\alpha_2=0.5+i(0.86)$,$\;$
$\beta_1=0.7+i(0.72)$, $\beta_2=0.44+i(0.9)$ and $\rho=\sigma =0.25$.
Thus for this case due to the interaction there is an exchange of periodic
intensity between the two modes of soliton $S_2$  with the suppression of the 
switching dynamics in soliton $S_1$. 
\section{CNLS with Coupled Cubic-Quintic Nonlinearity}
The model for ultra short optical pulse propagation in non-Kerr media
(in particular materials with high nonlinear coefficients even at moderate
optical intensities, for example, semiconductor dopped glasses, organic
polymers, etc.) can be obtained by expanding the induced polarization
vector in Maxwell's equation(1) as
\begin{equation}
\overrightarrow{P}(\overrightarrow{r},t)  =  \epsilon_0
[\chi^{(1)}.\overrightarrow{E}  +\chi^{(3)} \vdots 
\overrightarrow{E}\overrightarrow{E}\overrightarrow{E}
+\chi^{(5)}\dot{\dot{\vdots}} \overrightarrow{E}\overrightarrow{E}
\overrightarrow{E}\overrightarrow{E}\overrightarrow{E}+...], 
\end{equation}
where $\chi^{(5)}$ is the fifth order nonlinear susceptibility and
following the procedure as in the case of NLS equation. The resulting 
equation, in normalized form, describing the effects of quintic
nonlinearity on the ultra short optical soliton pulse propagation is [11],
\begin{equation}
iq_z +  q_{tt} + 2 |q|^2 q + \gamma |q|^4 q + i \gamma_1 q_{ttt} 
+ i \gamma_2 (|q|^2q)_t + i \gamma_3(|q|^4 q)_t = 0. 
\end{equation}
The generalization of equation(26) in order to include multimode effects
leads to the following coupled cubic-quintic NLS equation[11], after
neglecting third order terms,
\begin{eqnarray}
iq_{1z}+q_{1tt}+2(|q_1|^2+|q_2|^2)q_1+(\rho _1|q_1|^2 
+ \rho_2|q_2|^2)^2q_1
+2\rho _2[(\tau _1-\rho _1)|q_1|^2 \qquad\nonumber \\
+(\tau_2-\rho _2)|q_2|^2]\left| q_2\right| ^2q_1 
-2i[(\rho _1|q_1|^2 +\rho_2|q_2|^2)q_1]_t 
+2i(\rho _1q_1^{*}q_{1t} +\rho_2q_2^{*}q_{2t})q_1 &=&  0, 
\nonumber \\
iq_{2z}+q_{2tt}+2(|q_1|^2+|q_2|^2)q_2+(\tau _1|q_1|^2
+\tau_2|q_2|^2)^2q_2+2\tau _1[(\rho _1-\tau _1)|q_1|^2\qquad\nonumber\\
+(\rho_2-\tau _2) |q_2|^2]\left| q_1\right| ^2q_2
-2i[(\tau _1|q_1|^2+\tau_2|q_2|^2)q_2]_t
+2i(\tau _1q_1^{*}q_{1t}
+\tau _2q_2^{*}q_{2t})q_2 &= & 0,\nonumber\\
\end{eqnarray}
where $\rho_1, \rho_2, \tau_1 $ and $ \tau_2$ are real free parameters.
The integrable nature of the above equation can be studied by obtaining
the Lax pairs and conserved quantities for it. The Lax pair associated
with equation(27) is
{\small
\begin{eqnarray}
L= \left(
\begin{array}{ccc}
-i\lambda  & q_1                    & q_2 \\
-q_1^*     & \ -i\theta_{1t}+i\lambda & 0   \\
-q_2^*     & 0                      & -i\theta_{2t}+i\lambda  
\end{array} \right),\;\;  
M= \left(
\begin{array}{ccc}
-2i\lambda^2  & 
2\lambda q_1 &
2\lambda q_2
\\
+i(|q_1|^2   & 
+iq_{1t} &
+iq_{2t}
\\
+|q_2|^2)  & 
+\theta_{1t}q_1 &
+\theta_{2t}q_2
\\
\\
-2\lambda q_1^*  &
2i\lambda^2       &
-iq_1^*q_2
\\
+iq_{1t}^*  &
-i|q_1|^2       &
\\
-\theta_{1t}q_1^* &
-i\theta_{1z}        &
\\
\\
-2\lambda q_2^*  &
-iq_1q_2^*                                &
2i\lambda^2 
\\
+iq_{2t}^* &
                                 &
-i|q_2|^2 
\\
-\theta_{2t}q_2^* &
                               &
-i\theta_{2z}        
\end{array}
\right), 
\end{eqnarray}
}
where $\theta _1=\int^t_{-\infty} (\rho
_1|q_1|^2+\rho _2|q_2|^2)dt'$,
$\theta _2  =  \int^t_{-\infty} (\tau _1|q_1|^2+\tau
_2|q_2|^2)dt'$ and $\lambda$ is the spectral parameter. The 
compatability condition $L_z-M_t+[L,M]=0$ leads to equation(27).
Though the integrable system(27) possesses infinite number of conserved
quantities, only the lower ones are of physical importance. They
can be written as
\begin{eqnarray}
c_1 & = &-{1 \over 2i} \int_{-\infty}^{+\infty} dt (|q_1|^2+|q_2|^2), 
\nonumber \\ 
c_2 & = &- {i \over 4} \int_{-\infty}^{+\infty} dt [-i (q_1q_{1t}^*+ q_2q_{2t}^*)
+ \rho_1|q_1|^4 +\tau_2|q_2|^4\nonumber\\
& &  + (\rho_2+\tau_1)|q_1|^2 
|q_2|^2 ], \nonumber\\ 
c_3  &=& -{i \over 8} \int_{-\infty}^{+\infty} dt [(q_1q_{1tt}^*+ q_2q_{2tt}^*)
 + (|q_1|^2+|q_2|^2)^2+ i(|q_1|^2N_{1t}+|q_2|^2N_{2t})\nonumber\\
& & +2i(N_1q_1q_{1t}^* +N_2q_2q_{2t}^*)-(N_1^2|q_1|^2
 + N_2^2|q_2|^2)] , 
\end{eqnarray}
etc., where $N_1=\theta_{1t}= \rho_1|q_1|^2+ \rho_2|q_2|^2  $ and 
$ N_2=\theta_{2t}=\tau_1|q_1|^2+\tau_2|q_2^2|$.
Here $c_1$, $c_2$ and $c_3$
may be related to the number operator, angular momentum and the Hamiltonian
(energy) of the system(27) respectively. It is intriguing to note that
the fields $q_a$ and $q_a^*$, a=1,2 do not have canonical Poisson bracket
relations. However under the following nonultralocal Poisson bracket
structure,
\begin{eqnarray}
\{q_1(x), q^*_1(y)\}&=& \delta(x-y) 
 +i \rho_1 \epsilon (x-y)   q_1(x) 
q^*_1 (y),\nonumber\\
\{q_1(x), q_1 (y)\} &=&
 i \rho_1 \epsilon (y-x)   q_1(x) q_1 (y) ,\nonumber\\ 
\{q_1(x), q_2 (y)\}&= &- i (\rho_2 \theta (x-y)
 - \tau_1 \theta (y-x))q_1(x) q_2 (y),\nonumber\\
\{q_1(x), q_2^* (y)\}&= &i (\rho_2 \theta (x-y)
  - \tau_1 \theta (y-x))q_1(x) q_2^* (y),
\end{eqnarray}
etc., where $\epsilon (x) =\theta (x)-\theta (-x)$ is the sign function
defined through the step function, $\theta (x)= 1 $ for $x>0$, $\theta
(x)=0,$ for $x\le 0$, the integrals(29) become involutive.
The interrelation between equation(27) and the Manakov system allows 
us to construct the  soliton solution of equation(27) in terms of 
the known Manakov soliton solution.The fields of these two models are
related through a 
\begin{figure}[h]
\centerline{\epsfig{figure=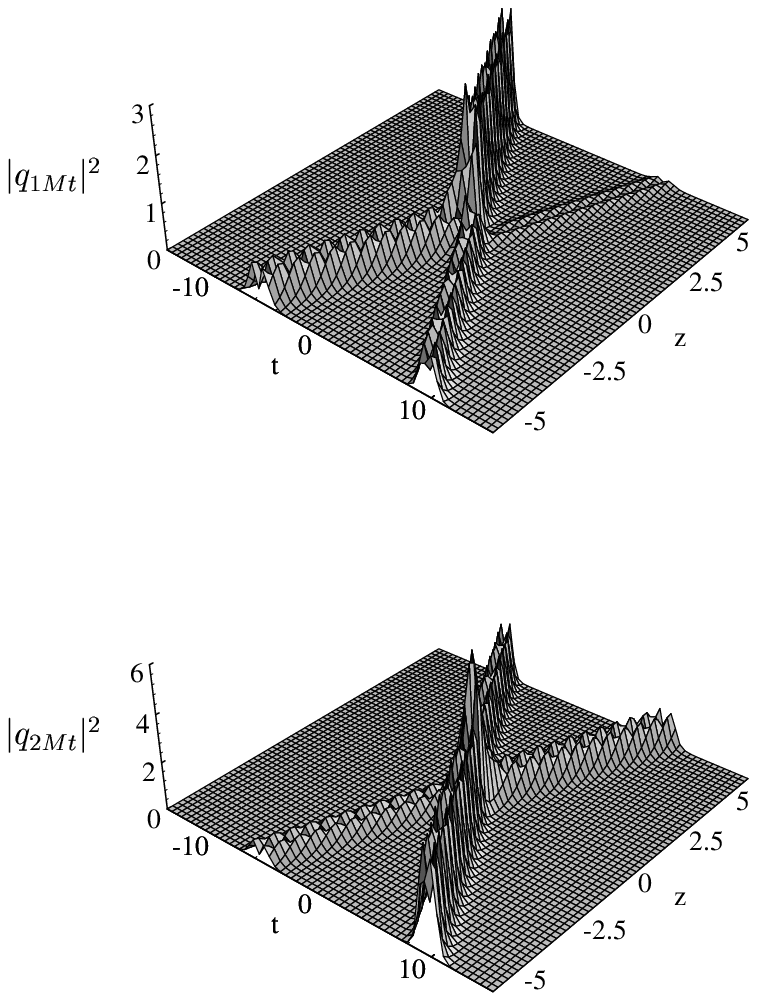, width=0.5\linewidth}
\epsfig{figure=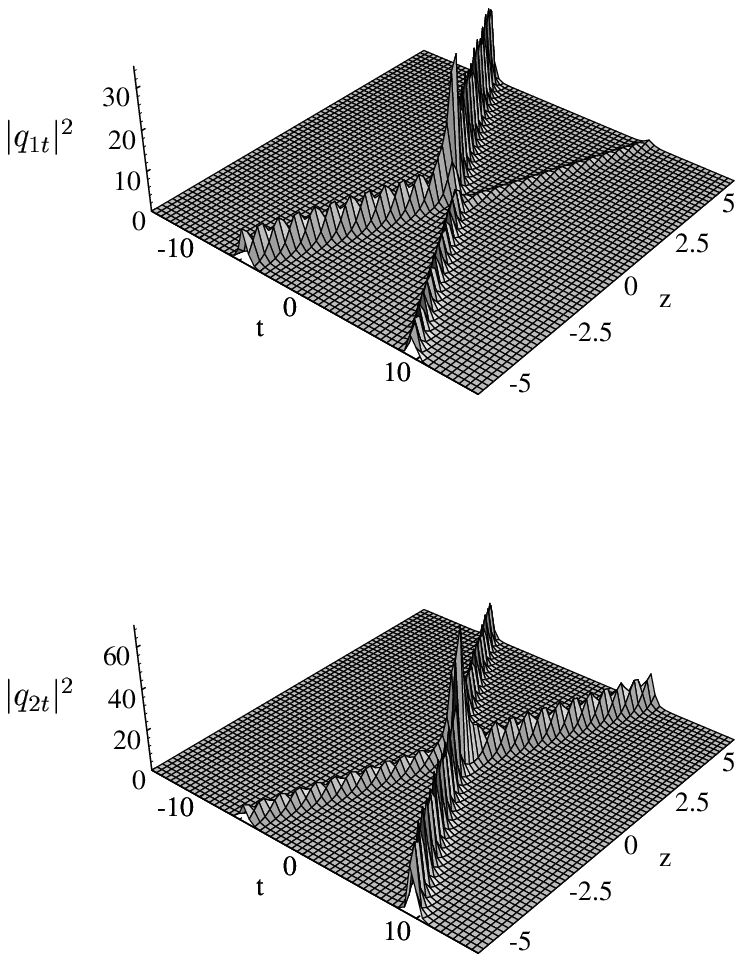, width=0.5\linewidth}}
\hspace{100pt}(a) \hspace{200pt}(b)
\caption{(a) Evolution of the intensity profiles $|q_{1Mt}|^2$ and
$|q_{2Mt}|^2$ of the two-soliton solution of the Manakov model. (b) Intensity
profiles $|q_{1t}|^2$ and $|q_{2t}|^2$ of the two-soliton solution of (27)
for $\rho_1=\rho_2=\tau_1=\tau_2=1.$ 
}
\label{3a}
\end{figure} 
nonlinear transformation for the dependent variables given by
\begin{equation}
q_{aM}=q_aexp(-i\theta_a),\;\;a=1,2
\end{equation}
and the subscrip M represents the Manakov model. Their Lax pairs are 
related through a local gauge transformation $L'=g^{-1}Lg-g^{-1}g_t$,
$M'=g^{-1}Mg-g^{-1}g_x$ where
\begin{eqnarray}
g = 
\left[
\begin{array}{ccc}
1 & 0 & 0\\
0 & exp(-i\theta_1) & 0\\
0 & 0 & exp(-i\theta_2)
\end{array}
\right].
\end{eqnarray}

Then the one-soliton solution takes the form
\begin{equation}
(q_{1}, q_{2} ) =
 (\alpha e^{i\delta_1 
\tanh (\nu (t-vz+\delta))},  
\beta e^{i\delta_2 \tanh (\nu (t-vz+\delta))})
 sech (\nu (t-vz +\delta))e^{i(\kappa t+\omega z)}
 \end{equation}
and the two-soliton solution can be obtained by substituting the
two-soliton solution of the Manakov model in the transformation(31).

The effect of the dependence of phase change on the real free parameters
$\rho_1$, $\rho_2$, $\tau_1$ and $\tau_2$ during collision can be revealed
by carrying out the asymptotic analysis of the two-soliton solution as
earlier. In this connection, we have taken the derivatives of the
$|q_j^{n\mp}|,$ n,j=1,2, since the effect of the phases are reflected only
in such terms rather than $|q_j^{n\mp}|^2$.

We have shown these effects for the parameter values $k_1=1.5+i(0.5),$
$k_2=2-i(0.7)$, $\alpha_1 = \beta_1 = \beta_2 = 1$ and $\alpha_2=\frac{
39+i80}{89}$, by comparing the plots of the intensity profiles in the absence
of the free parameters ($\rho_1$, $\rho_2$, $\tau_1$ and $\tau_2$) with those
in their presence. It is obvious from equation(31) that for the choice
of parameters $\rho_1$=$\rho_2$=$\tau_1$=$\tau_2$$=0$, the two-soliton solution of
equation(27) reduces to that of the Manakov system. Hence in Figure(3a)
we have plotted the intensity profiles of 
$|q_{1Mt}|^2$ and $|q_{2Mt}|^2$
for the above parametric values. In the figure there appears a splitting
in each of the asymptotic profiles before and after interaction.

After setting $\rho_1$=$\rho_2$=$\tau_1$=$\tau_2$$=1$, we can show that the 
splitting disappears as shown in Figure(3b). Thus in addition to the 
inelastic soliton collision, there is a change in the asymptotic form 
(suppression of splitting) of the intensity profiles which arises due
to the presence of the phase terms $\theta_a$ in the transformation(31). 
\section{General Soliton Perturbation}
As mentioned in section(2) the CNLS equation is in general a nonintegrable
one. We can apply the multiple scale perturbation theory to study such 
nonintegrable system with more general perturbations $R_1$ and $R_2$ as given
below,
\begin{eqnarray}
iq_{1z}+q_{1tt}+2\mu(|q_1|^2+|q_2|^2)q_1  = i\epsilon R_1(q_1,q_2,q_{1t},q_{2t}...)
,\nonumber\\
iq_{2z}+q_{2tt}+2\mu(|q_2|^2+|q_1|^2)q_2  = i\epsilon R_2(q_1,q_2,q_{1t},q_{2t}...).
\end{eqnarray}  
In the absence of the perturbation the above equation reduces to the Manakov
system. After expanding $q_i$ as 
\begin{equation}
q_i=q_i^{(0)}(k_{iR},k_{iI},A_i^{j\pm},\Phi_i^{j\pm})+\epsilon q_i^{(1)}
+\epsilon^2 q_i^{(2)}+...\; \mbox{i, j=1,2,}
\end{equation}
where $A_i^{j\pm}$ and $\Phi_i^{j\pm}$ are the real and imaginary parts of the 
polarization vectors of the two-solitons respectively, in the asymptotic limits
$z\rightarrow \pm\infty$. Substitution of this in equation(34), it is trivial to
verify that $q_i^{(0)}$ is nothing but the two-soliton solution of the Manakov
system. Asymptotically this two-soliton solution of the Manakov model given by
equation(17) can be written as a combination of two one-solitons as 
$z\rightarrow \pm\infty$
\begin{eqnarray}
\left(
\begin{array}{c}
q_1\\
q_2 
\end{array}
\right)
\rightarrow
\left(
\begin{array}{c}
q_1^{1\pm}+q_1^{2\pm}\\
q_2^{1\pm}+q_2^{2\pm}
\end{array}
\right). 
\end{eqnarray}
By using a direct perturbational approach [12] the evolution equations for the
soliton parameters in the presence of perturbations can be obtained. Applying this
method to CNLS equation(10) one can easily check that there occurs transmission
and reflection scenarios during collision with a sensitive dependence of the
collision outcome on the $\;$cross phase modulation coefficient and initial soliton
parameters. As a special case by taking $R_i=(-\gamma q_i+\bar\beta|q_j|^2q_i
+\frac{\delta}{2}q_{itt})$ where i,j=1,2 we have studied the effect of 
perturbations such as fiber loss, dispersion gain and nonlinear gain. The
importance of taking such a form for perturbation is that in the absence of
$q_1$ and $q_2$ equation(34) along with the above specified form of $R_1$ and $R_2$
reduces to the Ginzburg-Landau equation governing pulse propagation in fiber
amplifiers [1]. Since we have taken the asymptotic form of the more general
two-soliton solution as the zeroth order solution, the collision properties of the
system(34) can be directly studied by comparing the evolution of the soliton
parameters in the limits $z\rightarrow -\infty$ and $z\rightarrow \infty$.
We expect that this may lead to some novel results such that collision among the
solitons may be used to compensate fiber loss experienced by the interacting
coupled one-solitons corresponding to any one of the polarized modes.
\section{Acknowledgement}
This work is supported by the Department of Science and Technology, Government of
India in the form of a Research Project.

\end{document}